\documentclass[mathleft
]{an}
\usepackage{graphicx}
\usepackage{times}
\usepackage{natbib}
\bibpunct{(}{)}{;}{a}{}{,}
\begin{document}

\Pagespan{1}{}
\Yearpublication{..}%
\Yearsubmission{2012}%
\Month{08}%
\Volume{..}%
\Issue{..}%

\title{Fundamental parameters of FR~II radio galaxies and their impact on groups and clusters' environments}

\author{A.D. Kapi\'{n}ska\inst{1}\fnmsep\thanks{Corresponding author:  \email{anna.kapinska@port.ac.uk}\newline}
\and  P. Uttley\inst{2}
}
\titlerunning{FR~II radio galaxy impact on clusters' environments}
\authorrunning{A.D. Kapi\'{n}ska \& P. Uttley}
\institute{
Institute of Cosmology \& Gravitation, University of Portsmouth, Burnaby Road, PO1 3FX Portsmouth, U.K.
\and 
Astronomical Institute `Anton Pannekoek', University of Amsterdam, Science Park 904, 1098 XH Amsterdam, Netherlands}

\received{24 Aug 2012}
\accepted{.... }
\publonline{later}


\abstract{%
Radio galaxies are among the largest and most powerful single objects known and are found at variety of redshifts, hence they are believed to have had a significant impact on the evolving Universe. Their relativistic jets inject considerable amounts of energy into the environments in which the sources reside; thus the knowledge of the fundamental properties (such as kinetic luminosities, lifetimes and ambient gas densities) of these sources is crucial for understanding AGN feedback in galaxy clusters. In this work, we explore the intrinsic and extrinsic fundamental properties of Fanaroff-Riley II (FR~II) objects through the construction of multidimensional Monte Carlo simulations which use complete, flux limited radio catalogues and semi-analytical models of FR~IIs' time evolution to create artificial samples of radio galaxies. This method allows us to set better limits on the confidence intervals of the intrinsic and extrinsic fundamental parameters and to investigate the total energy produced and injected to the clusters' environments by populations of FR~IIs at various cosmological epochs ($0.0<z<2.0$). We find the latter estimates to be strikingly robust despite the strong degeneracy between the fundamental parameters -- such a result points to a conclusive indicator of the scale of AGN feedback in clusters of galaxies. 
}

\maketitle

\section{Introduction}

Radio galaxies and quasars are among the largest and most powerful single objects known. Since they often reside in groups and clusters of galaxies and are found at variety of redshifts these radio sources are believed to have had significant impact on the evolving Universe \cite[e.g.][]{1979MNRAS.189..433L, 1997ApJ...476..489, 2001ApJ...560L.115G, 2004MNRAS.355L...9R}. Their relativistic jets inject considerable amounts of energy into the ambient medium which surrounds active galactic nucleus (AGN); hence these sources are invaluable laboratories in studies of intergalactic and intracluster medium, and in studies of the evolution of their hosts \cite[e.g.][]{2005ApJ...635...13}.

Although the original models of formation and evolution of galaxy clusters \cite[e.g.][]{1977ApJ...215..723C, 1977MNRAS.180..479F}, which often are the habitats of radio galaxies,  predicted self-similar growth of the structures, it soon became apparent that they over-estimate cooling rates of the evolving clusters \cite[the cooling flow problem; see][and references therein]{2007ARA&A..45..117M}; if only gravitational heating is included in the models, too many baryons cool to form stars and galaxies become too luminous. Moreover, scaling relations of galaxy clusters' properties (such as mass, temperature, X-ray luminosity) predicted by these self-similar models have also been found to diverge from the observationally derived relations \cite[similarity breaking; e.g.][]{1998ApJ...504...27M}. Mechanisms which have been proposed to explain these problems, referred to as non-gravitational heating, include supernovae and AGN feedback, sub-cluster merging, and thermal conduction among others \cite[e.g.][]{1989ApJ...338..761R, 1999Natur.397..135P, 2001ApJ...563...95M, 2004MNRAS.348.1105O}. Here  we focus on the AGN feedback in galaxy groups and clusters, and on the impact of FR~II \cite[][]{1974MNRAS.267...31} radio galaxies in particular. The main effects an AGN exerts on its surroundings include removing cold low entropy gas from the centres of clusters and groups to the outskirts of these structures. Through the work done on the environments by the AGN jets the excess energy is transferred to the ambient gas -- this will cause an increase in the ambient temperature as well as in the cluster gas entropy.

In these proceedings, we investigate the total energy injected by FR~II radio galaxies and radio-loud quasars into their environments. In our recent paper \cite[hereafter KUK12]{kuk12} we presented a new numerical method which allows one to explore the most likely fundamental parameters that define whole populations of FR~II radio sources, i.e. distributions of kinetic luminosities $Q$, their lifetimes $t_{\rm max}$ and central gas densities $\rho_{o}$, as well as the number densities of FR~II radio sources, at various cosmological epochs -- these have long been a sought-after piece of information in the study of radio galaxies and quasars \cite[e.g.][]{1991Natur.349..138R, 1995ApJ...454..580D, 2007ApJ...658..217B}. However, KUK12 showed that these fundamental parameters are highly degenerate and to disentangle information independent constraints on some of the parameters are required. These results led to questions which we attempt to answer here: {\it Is the scale of the impact of FR~II radio sources on clusters' environments robust despite the degeneracy?  Or, can the degeneracies be broken? }

The article is structured as follows. The method used and the results from KUK12, which this paper is based on, are briefly presented in \S\ref{sec:method}. The total energy produced by FR~II radio galaxies and quasars is reported in \S\ref{sec:totalenergy-res:totalinjected} and discussed in \S\ref{sec:discussion}. The summary is given in \S\ref{sec:summary}. A flat Universe with the Hubble constant of $H_0 = 71$ km~s$^{-1}$~Mpc$^{-1}$, and $\Omega_{\rm M} = 0.7$ and $\Omega_{\Lambda} = 0.3$ is assumed throughout these proceedings.


\begin{table*}
  \centering
  \caption{Fundamental parameters defining populations of FR~II radio sources which were chosen for the study. For details see \S\protect\ref{sec:method}.}
  \label{tab:totalenergy-datasets}
  \begin{tabular}{cccccccc}
    \hline
    \hline
\\
    Data set & $Q_{\rm B}(z_0)$  [log(W)]  & $n_{\rm q}$& $\alpha_{\rm s}$ & $\rho_{\rm m}(z_0)$  [log(kg m$^{-3}$)] & $n_{\rm r}$ & $t_{\rm max_{\rm m}}(z_0)$ [log(yr)] & $n_{\rm t}$ \\
\\	
    \hline
    $1a$ & 38.0  & \phantom{0}8.0 & 0.6 & $-23.3$ & \phantom{1}4.0           & 7.8 & $-2.5$\\
    $1b$ & 38.0  &           10.5 & 0.6 & $-23.0$ & \phantom{1}0\phantom{.0} & 7.8 & $-4.0$\\
    $2$  & 39.2  & \phantom{0}4.0 & 0.4 & $-25.1$ &           10.5           & 6.9 & \phantom{$-$}$0$\phantom{.0}\\
    $3$  & 37.4  & \phantom{0}8.0 & 0.6 & $-21.5$ & \phantom{1}5.5           & 8.7 & $-3.0$\\
    \hline
    \hline
  \end{tabular}
\end{table*}


\section{Methods}
\label{sec:method}

The method which we use in our investigations is based on a construction of multidimensional Monte Carlo simulations, which use semi-analytical models of FR~IIs' time evolution to generate artificial samples of radio sources. We use the theoretical model of \cite{ka97} and \cite{kda97} and wide ranges of searched values of the fundamental parameters with which the models are used in this work. Once the artificial samples are generated one can easily find their radio luminosity functions \cite[RLFs; ][]{1968ApJ...151..393S}; these are further compared to the observed RLFs of FR~II sources at various redshifts. Through the maximum likelihood method we found the best fitting FR~IIs' fundamental parameters, and their corresponding confidence intervals. We use complete, flux-limited radio samples, which in this work include the Third Cambridge Revised Revised Catalogue \cite[3CRR; ][]{1983MNRAS.204..151L} and radio sample constructed by \cite{1999MNRAS.310..223B}. For all relevant details and assumptions adopted the reader is referred to KUK12. 

As reported in KUK12, we found that the best fitting fundamental parameters span wide ranges of possible values, extending over a few orders of magnitude. Although the estimated kinetic luminosities fall within the expected range of $10^{37}-10^{41}$~W for FR~II radio sources \cite[e.g.][]{1991Natur.349..138R}, the location of the kinetic luminosity break ($Q_{\rm B}$) of the Schechter (\citeyear{1976ApJ...203..297S}) function (which was assumed to describe distributions of kinetic luminosities) was found to be uncertain due to strong degeneracies between the parameters. For instance, in the local Universe, when allowing radio sources to have relatively short lifetimes ($10^7$~yr) low gas density environments ($\sim10^{-25}$~kg~m$^{-3}$) and breaks at high kinetic luminosities ($10^{39}-10^{40}$~W) are required in order to reconstruct the observed RLFs. Whereas for the longest lifetimes ($10^9$~yr) of these sources much denser environments ($\sim10^{-21}$~kg~m$^{-3}$) and lower $Q_{\rm B}$ ($\sim 10^{38}$~W) are necessary. 
Moreover, these parameters have also been found to evolve with cosmological epoch, where we assumed that the parameters change with redshift according to $Q_{\rm B}(z) = Q_{\rm B}(z_0)(1+z)^{n_{\rm q}}$ for the kinetic luminosity break, and $t_{\rm max_{m}}(z) = t_{\rm max_{\rm m}}(z_0)(1+z)^{n_{\rm t}}$ and $\rho_{\rm m}(z) = \rho_{\rm m}(z_0)(1+z)^{n_{\rm r}}$ for distributions of maximum lifetimes and central gas densities respectively, and where $z_0$ denotes $z=0$. A strong degeneracy has been observed to occur also in $n_{\rm q}, n_{\rm t}$ and $n_{\rm r}$ parameters.

 
\begin{figure} 
\includegraphics[width=62mm, angle=270]{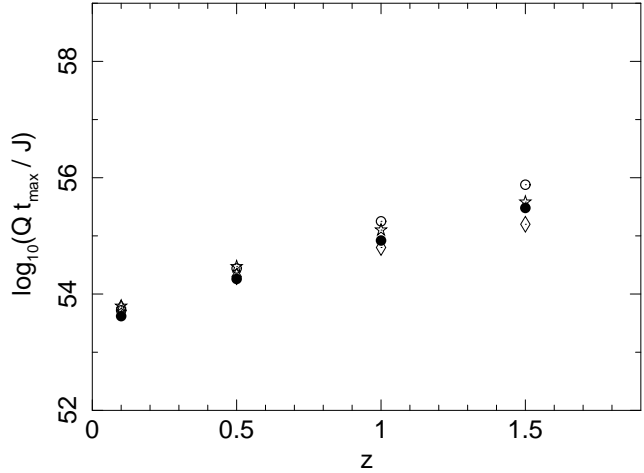} 
\caption{Total energy produced by FR~II radio galaxies during their lifetimes at different comic epochs. Data sets (consult Table \protect\ref{tab:totalenergy-datasets}) are marked as different data points as follows: data set $1a$ is marked as filled circles, data set $1b$ as open circles, data set $2$ as diamonds, and data set $3$ as stars. For details see \S\protect\ref{sec:method}.}
\label{rys:total-injected-energy} 
\end{figure} 


Despite these degeneracies one may attempt to estimate the total energy ($U_{\rm tot}$) produced by a radio source during its lifetime, where 
\begin{equation}
U_{\rm tot} = \int^{t_{\rm max}}_0 Q(t) {\rm d}t.
\label{eqn:utot}
\end{equation}
In this preliminary analysis we selected four data sets  to investigate estimates on the total energy. The data sets are within the 90\% confidence intervals of the best fits as found in KUK12. We consider the best fit found (data set $1b$) and a data set in which the cosmological evolution of central gas densities is set to $n_{\rm r}=4.0$ (discussed in KUK12, here data set $1a$). In addition, we draw two data sets ($2$ and $3$) that represent the two extreme cases discussed above (that is cases when the maximum lifetime is set to the shortest and longest one found within 90\% confidence intervals of KUK12 results, see also Table~\ref{tab:totalenergy-datasets}). Furthermore, it is assumed that the break kinetic luminosities of each data set represent the typical luminosities of radio sources in their respective populations, and that radio sources at the break kinetic luminosities have maximum lifetimes and central gas densities typical for their populations. Based on these assumptions and since the kinetic luminosity is assumed to be constant over the radio source's life \cite[][]{ka97}, Eqn.~\ref{eqn:utot} simplifies to 

\begin{equation}
U_{\rm tot} = Q_{\rm B} t_{\rm max_{m}}.
\label{eqn:simplified}
\end{equation}

We investigate the total power and total energy produced by FR~IIs' populations at four redshift steps, up to $z=1.5$.

\section{Results}
\label{sec:totalenergy-res:totalinjected}

The total energy produced by FR~II radio sources for each data set and redshift considered is presented in Fig.~\ref{rys:total-injected-energy}. Interestingly, the results are alike for each data set. As can be deduced from KUK12 results, the higher the kinetic luminosity, the shorter the lifetime of the sources is required, and hence, as we explicitly show here, the product of these degenerate results ($U_{\rm tot}$) is robust.

Furthermore, the total power produced by FR~IIs seem to increase with redshift. Kinetic luminosities and maximum lifetimes of FR~II radio sources undergo contrasting cosmological evolution, where kinetic luminosities of the sources increase with redshift, while their maximum lifetimes decrease with cosmic epoch. Since the cosmological evolution of the kinetic luminosity breaks and maximum lifetimes of the radio sources is known in our calculations (Tab.~\ref{tab:totalenergy-datasets}) the plausible redshift evolution of $U_{\rm tot}$ can be estimated. We note that Eqn.~\ref{eqn:simplified} can be expressed with redshift dependence, i.e. 

\begin{equation}
U_{\rm tot}(z) = Q_{\rm B}(z_0) t_{\rm max_{m}}(z_0) (1+z)^{n_{\rm q} + n_{\rm t}},
\end{equation}
which leads to $U_{\rm tot} \propto (1+z)^n$ with  $n\in [4.0; 6.5]$ in our current preliminary calculations. This result needs to be confirmed with statistics on full populations, however.

\section{Discussion}
\label{sec:discussion}

We find estimates of the total energy produced by populations of FR~II radio sources to be remarkably robust despite the strong degeneracy between the fundamental parameters. Such a result points to a compelling indicator of the scale of AGN feedback in groups and clusters of galaxies. 

The estimated kinetic luminosities and available energy of FR~II radio sources suggest that radio galaxies are able to quench cooling flows at least for the duration of their lifetimes \cite[Fig.~\ref{rys:total-injected-energy}; cf.][and references therein]{2012AdAst2012E...6G}. It has been suggested before that AGNs with powerful radio outbursts may balance out the clusters' cooling flows \cite[e.g.][]{2004ApJ...607..800B, 2006ApJ...652..216R}. The range of possible values of kinetic luminosities (consult KUK12) may suggest that the estimates from our degenerate results of quenching cooling flows with radio galaxies are less robust than those of total energy produced (these proceedings). However, the more powerful outbursts are found to last for shorter time than the less powerful ones (Tab.~\ref{tab:totalenergy-datasets}). If activity in radio galaxies is recurrent, galaxy clusters hosting such radio sources would undergo alternate periods of cooling and heating, but the recurrence timescales may be related to the power of the radio source involved, where the weaker outbursts would be expected to occur more often.

It is important to note here the difference between morphological classes of radio galaxies, that is FR~I and FR~II, and the possible difference in their impact on clusters' environments. As pointed out by \cite{2011ApJ...738..155M} linear sizes of typical radio galaxies significantly exceed clusters' cooling radii \cite[$r_{\rm cool}\sim100$~kpc; e.g.][]{2007MNRAS.379..894B}. This is especially relevant in the case of FR~II sources since due to their large linear extent and their strong jet head driven shocks one must expect most of the energy to be deposited beyond the cooling radius \cite[e.g.][]{2007ApJ...660.1118G,2011ApJ...738..155M}. 
At lower redshifts ($z<1.0$) it is the FR~I sub-population that dominates investigations carried out with X-ray observations  \cite[e.g.][]{2010ApJ...710..743D}. Due to the difference in interaction with the ambient medium of the two morphological classes, it has been suggested that FR~Is mainly contribute to heating of cluster cores, while FR~IIs contribute to overall AGN feedback, especially at higher redshifts where they are expected to be much more numerous, and may be more relevant in the problem of similarity breaking \cite[e.g.][]{2010ApJ...710..743D}. 
Our work is currently limited to radio sources of FR~II morphology only, due to the availability of the theoretical models of radio source growth with time. In contrast to turbulent structures of FR~Is, the idealised, and often regular, large scale radio structures of FR~IIs make them relatively simple systems to model. Although there have been attempts to develop models of FR~I time evolution \cite[e.g.][]{1995ApJS..101...29B, 2010ApJ...713..398L} they are still not general enough to be applied to whole populations of these radio sources. Once such models are developed, our analysis can easily be extended to include populations of FR~I sources.

Finally, despite the striking robustness of the total energy estimates, one should consider possible methods for breaking the degeneracy between the fundamental parameters; we will shortly list them here. As pointed out by \cite{2002MNRAS.332..729C} the AGN should be finely tuned and neither too much nor too little heating can be accepted. Since the levels of required energy in order to converge observations and theoretical models are determined \cite[see e.g.][and references therein]{2001ApJ...546...63T} these estimates can be used as a tool for breaking degeneracy in the KUK12 results. Additionally, X-ray observations can provide one with estimates of ambient gas densities of FR~II radio sources at least in the local Universe, which again can be used to help in breaking the degeneracies.

\section{Summary}
\label{sec:summary}

We extend the work presented in KUK12 and investigate the total energy produced by populations of FR~II radio sources during their lifetimes, and the sources' impact on their environments, at various cosmological epochs. In these preliminary results we report that this total energy seems to be remarkably robust despite strong degeneracies between the fundamental parameters defining populations of FR~II sources (their kinetic luminosities, lifetimes and gas densities of the environments). These results hold for all redshifts considered. Moreover, we find that the total energy produced by FR~II radio sources increase with redshift. 
Such results point to a compelling indicator of the scale of AGN feedback in groups and clusters  of galaxies across cosmological times.

\acknowledgements
The authors thank the anonymous referee for comments on the manuscript. This work made use of the Iridis Compute Cluster maintained by the University of Southampton, U.K. ADK acknowledges financial support from the Leverhulme Trust received  during the time this work has been conducted.


\end{document}